\documentclass[letter,12pt]{article}
\usepackage[T1]{fontenc}
\usepackage{amsmath}
\usepackage{amssymb,amsfonts,textcomp}
\usepackage{array}
\usepackage{hhline}
\usepackage[pdftex]{graphicx}
\usepackage{mathrsfs}
\usepackage[latin1]{inputenc} % En caso de usar tildes, Ã± y otros caracteres especiales
\usepackage{amsfonts,amsthm}

\theoremstyle{definition}

\numberwithin{equation}{section}
\newtheorem{teorem*}[thrm]{Theorem}
\usepackage{color}

\evensidemargin = \oddsidemargin
\newcommand{\nada}[1]{}

\begin{document}

\title{Using generalized logistics regression to forecast  population infected by Covid-19}

\author{ Villalobos Arias, Mario Alberto\thanks{Universidad de Costa Rica, CIMPA y Escuela de Matemática, San José, Costa Rica, mario.villalobos@ucr.ac.cr\newline Instituto tecnologico de Costa Rica, Escuela de Matemática, Cartago, Costa Rica, marvillalobos@itcr.ac.cr} 
}

\bigskip

\maketitle

\begin{abstract}
In this work, a proposal to forecast the populations using generalized logistics regression curve fitting is presented.
This type of curve is used to study population growth, in this case population of people infected with the Covid-19 virus; and it can also be used to approximate the survival curve used in actuarial and similar studies.

% \centerline{{\bf  Resumen}}

% En este trabajos se presenta una propuesta para la estimación de la poblaciones usando ajuste de curvas del tipo logística.
% Este tipo de curvas se utilizan para el estudio de crecimiento de poblaciones, en este casos población de personas infectadas por el virus Covid-19; y también se puede utilizar para aproximar la curva de supervivencia que se utiliza en estudios actuariales y otras similares.
\end{abstract}


% {\bf palabras clave:}Heurísticas de optimización, regresión logística generalizada, ajuste de curvas, covid-19.

\section{Introduction} 
\begin{figure}[ht]
\centerline{\includegraphics[width=9cm,height=6cm]{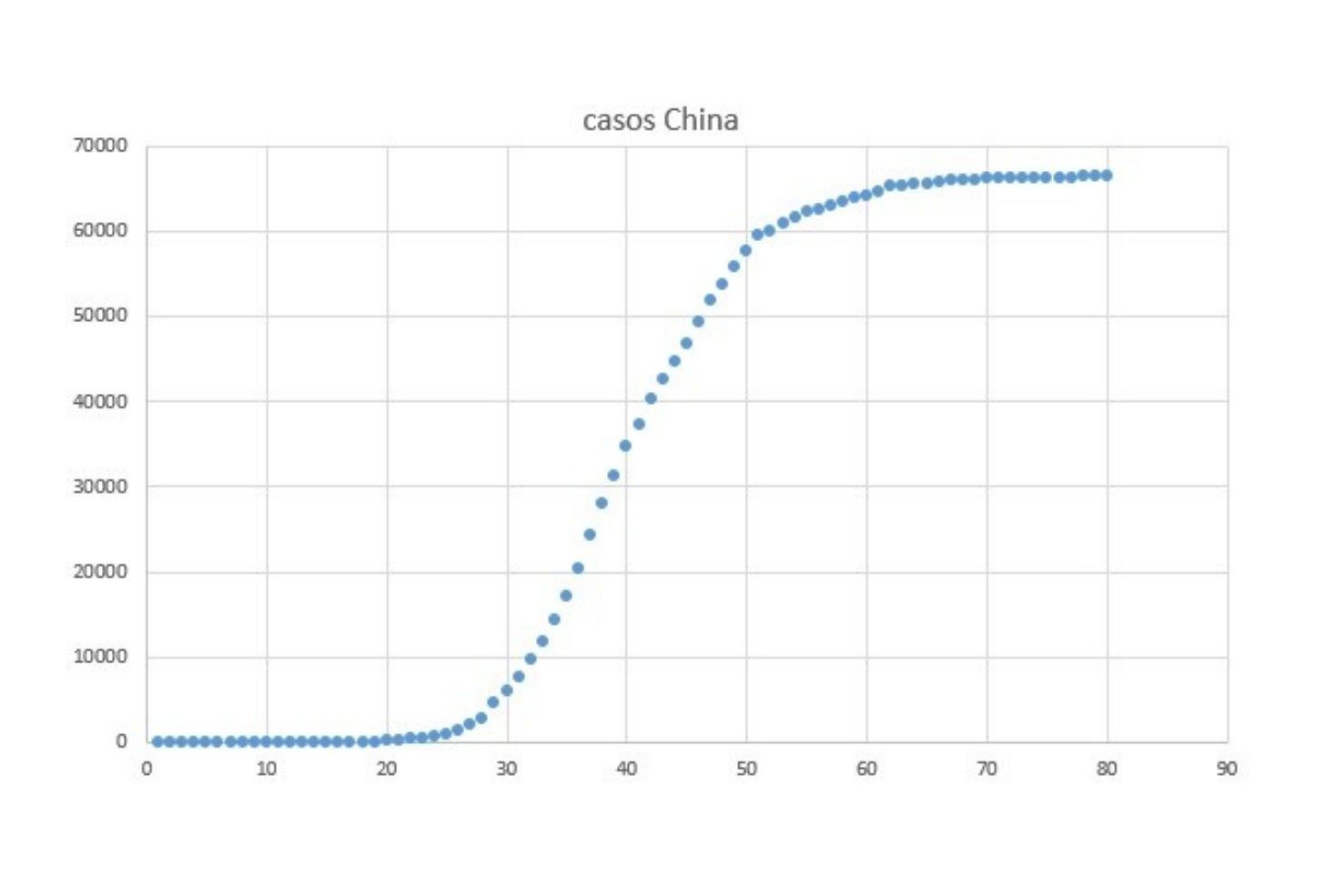}}
\caption{casos totales de contagiados en China}
\label{datosChina}
\end{figure}

Population growth curves follow the well-known logistic behavior as shown in the figure \ref{datosChina}.

A model that is used to fit population curves is the Logistic one, which uses the following equation
\begin{equation}
P(t) \ = \ \frac{1}{1+e^{-at+b}}
\label{logistica}
\end{equation}

As you can see this model has several problems among them that the data is in $ [0,1] $, and it is not flexible, the advantage is that this model is that an approximation of the optimal solution can be obtained by transforming the data and use linear regression.

\begin{figure}[ht]
\centerline{\includegraphics[width=9cm,height=6cm]{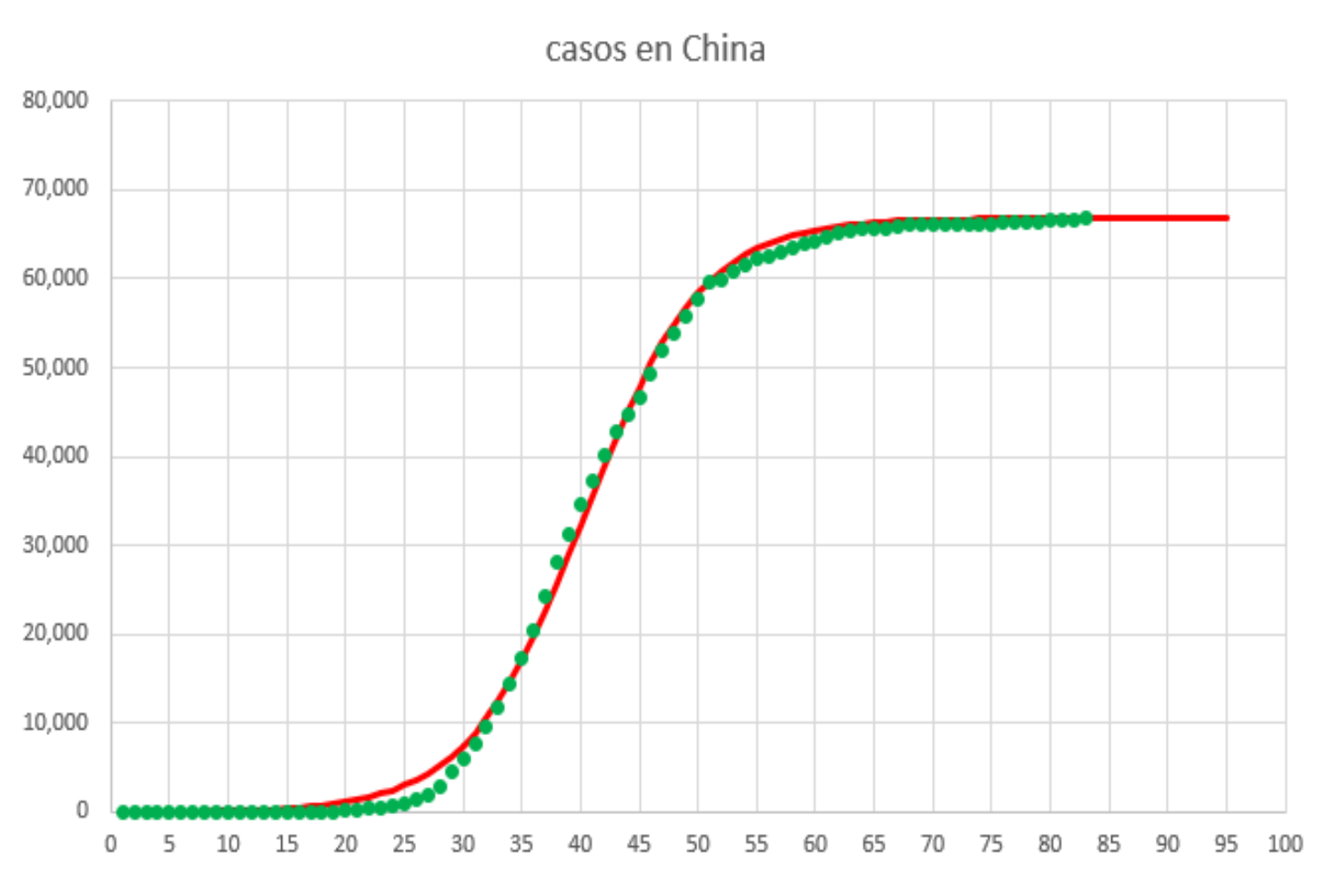}}
\caption{fit with logistic regression, total of infected in China}
\label{logChina}
\end{figure}

As seen in the graph \ref{logChina} applying logistic regression and dividing all the data by the maximum value (66818) a curve that fits well is obtained, with a $ R^2 = 0.99955$, which is very good , from the statistical point of view, but as seen in the figure there are many values that do not fit very well in the curves and it is not good for prediction, as we will see later.

\section{SIR model}

The classic version for the study of epidemics is the SIR model in which the population is divided into three groups: the susceptible, the infected and the recovered (SIR), this in the simplest case and that the population is changing from susceptible infected and then recovered.

The decrease in susceptibles is assumed to be proportional to the number of infected multiplied by the same number of susceptibles.
The change in the recovered is equal to a certain percentage of the infected and finally the number of infected is going to change increasing by the susceptibles that are infected and then we take away the amount of the infected that is recovered. In this way the following equations are obtained:

$$
{\displaystyle {\begin{aligned}&{\frac {dS}{dt}}=-{\frac {\beta IS}{N}},\\[6pt]&{\frac {dI}{dt}}={\frac {\beta IS}{N}}-\gamma I,\\[6pt]&{\frac {dR}{dt}}=\gamma I,\end{aligned}}}
$$
with
$$
{\displaystyle {\frac {dS}{dt}}+{\frac {dI}{dt}}+{\frac {dR}{dt}}=0,}
$$
which gives
$$
{\displaystyle S(t)+I(t)+R(t)={\text{constant}}=N,}
$$

The problem with the SIR model is that, as we see, the three parameters that are in the differential equations are needed, but with few data or with a few days it is very difficult to determine those parameters.

\section{Generalized Logistic Regression}

Therefore, the use of a slightly more complex model to the logistic regression and that is easier to determine the parameters than the SIR are proposed.

A first version is:
\begin{equation}
P(t) \ = \ \frac{M}{1+e^{-at+b}}
\label{logGenM}
\end{equation}
In this one more parameter is added, to be determined, which is the population limit $ M $, since with this modification it cannot be solved by transformation and linear regression.
To solve this problem, nonlinear optimization techniques must be used, it also has the stiffness problem, that is, it does not fit sufficiently to certain parts of the curve, this model can NOT be used for prediction, but from the point of adjustment it does not it is as accurate as we will see later.
 
To make the curve more flexible, an extra parameter $ \alpha $ is added as follows:
\begin{equation}
P(t) \ = \ \frac{M}{\big(1+e^{-at+b}\big)^{\alpha}}.
\label{logGen}
\end{equation}

This parameter $ \alpha $ adds flexibility in fitting the curve, remember the graphs of $y=x^{1/3}$, $y=x^{1/2}$, $y=x$,$y=x^2$,$y=x^3$.

For this function the inflection point is obtained when 
$$
P''(t) \ = \ a^2 c e^{a t + b} (e^{a t + b} + 1)^{-c - 2} (c e^{a t + b} - 1) \ = \ 0
$$
so the inflection point is obtained for
$$
t= -\frac{\ln (\alpha)+b}a
$$

This is the same model known as the Richards curve that is used to model population growth. 
\begin{equation}
Y(t)\ = \ A+{K-A \over (C+Qe^{{-Bt}})^{{1/\nu }}}
\end{equation}
Note that with some calculations and $ A = 0 $ equality with generalized logistics is given

The case:
$$
Y(t)={K \over (1+Qe^{{-\alpha \nu (t-t_{0})}})^{{1/\nu }}}
$$
which is a solution of the differential equation:
$$
Y^{{\prime }}(t)=\alpha \left(1-\left({\frac  {Y}{K}}\right)^{{\nu }}\right)Y
$$
\subsection{Gompertz Function }
It owes its name to Benjamin Gompertz, the first to work in this type of function is a particular case of Richards, and has the following equation:  
$$
{\displaystyle G(t)=a\mathrm {e} ^{-b\mathrm {e} ^{-ct}},}
$$
Furthermore, its second derivative is:
$$
G''(t) \ = \ b c^2 e^{b e^{c x} + c x} (1 + b e^{c x}) \ = \ 0
$$
which gives us the tipping point is reached in $t= -log(-b))/c$
what in the case of epidemics tells us at what point the growth of daily cases will start to decrease.

This is a simpler function since it only has 3 parameters instead of the LG, which has 4, and therefore it will have less local optimums.

\subsection{Other versions}
Other more complex versions are: 
\begin{equation}
P(t) \ = \ \frac{M + ct}{\big(1+e^{-at+b}\big)^{\alpha}}.
\label{logGenrecta}
\end{equation}

At the time of this writing, the South Korea covid-19 data has to be fitted with a curve like this.

And the next one that was used to adjust the survival curves
\begin{equation}
P(t) \ = \ \frac{M}{\big(1+e^{-at^2+bt+c}\big)^{\alpha}},
\label{logGenCuad}
\end{equation}

\section{Hypothesis and proposal}

The proposal of this work is, first, to use the generalized logistic curve or the Gompertz curve to make an adjustment of the data in which the curve is almost complete, for example data from China or South Korea, (March, 30)

On the other hand, when all the data are available, there is a generalized logistic regression or Gompertz type curve that fits the data.

So the hypothesis here is:
If we have the lower part of the curve, that is, the first values of the curve, we can obtain the parameters of the curve, and obtain the complete curve. And with this it can be used to predict population growth, in this case, for example, the total number of cases by covid-19 in a country or region and when the inflection point is reached, that is, when that the number of daily cases begins to decrease.

More specifically, if you have the data for about 20, 30 or 35 days, the question is:
 
Can we determine the curve parameters that fit the complete data?

If this will be achieved as we see, we would have a way of predicting the behavior of population growth with just having a few days, that is, we can determine the parameters of the curve with a few days, which is very difficult with the SIR model to determine the parameters of the differential equations.

\section{Curve Fitting}

For this work, the data provided by the ``European Centre for Disease Prevention and Control'' are being used, on the website to download the daily data, see \cite{datos}.

The first thing that is presented is to verify, as is known, that the LG curve fits the data from covid-19 very well. In the case of China and South Korea, which are the ones that have the almost complete curve.

Using a nonlinear optimization algorithm, the following results are obtained.

\subsection{China: original data}
\label{ChinaO}
\begin{figure}[ht]
\centerline{\includegraphics[width=9cm,height=6cm]{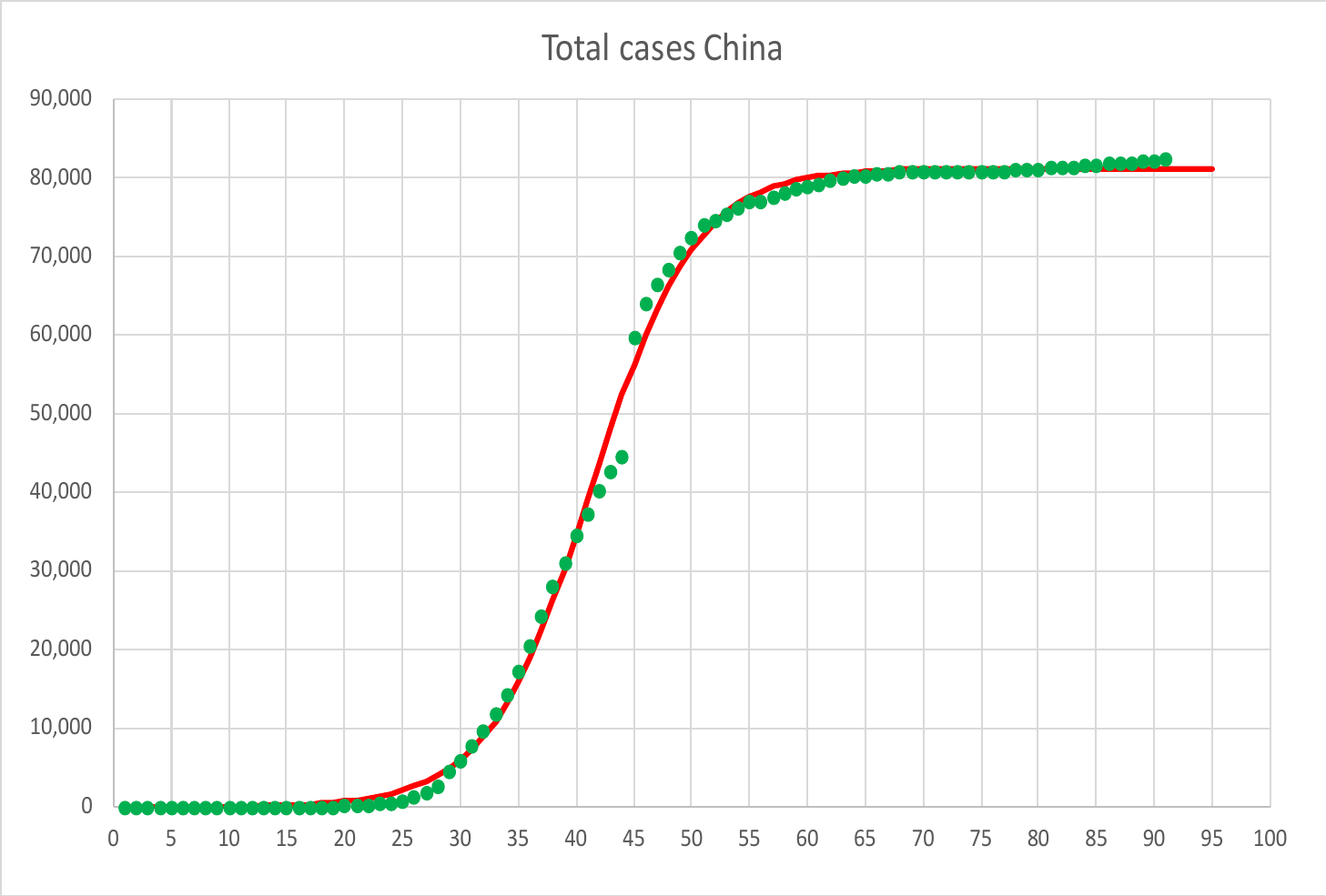}}
\caption{fiting with logistic regression, total of infected in China}
\label{ChinaOri}
\end{figure}

\begin{figure}[ht]
\centerline{\includegraphics[width=9cm,height=6cm]{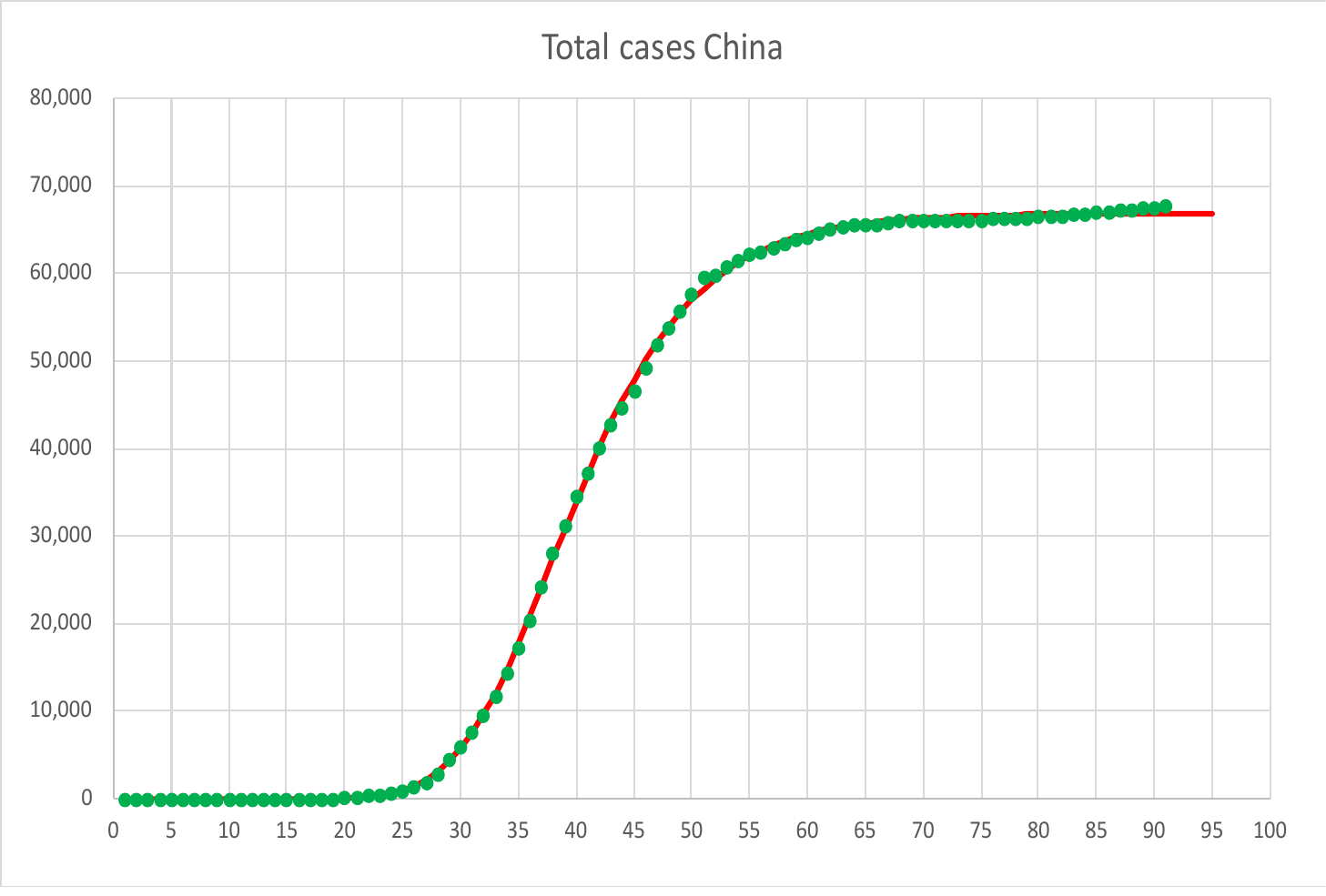}}
\caption{Fit with generalized logistic regression, China {\bf corrected data}}
\label{ChinaCorr}
\end{figure}

For the China data, the following parameters are obtained, and the graph in figure \ref{ChinaOri}.

$$
\begin{array}{ccccc}
fecha &  	M &  	a &  	b &  		\alpha \\\hline 	
31/03/2020 &  	81149 &  	-0.2348563&  	9.89996092 &  	0.8809852 \\\hline 	
\end{array}
$$
Con $R^2=	0.998878.$
As it is observed, a very good approximation is obtained, but as we know and it is seen in the figure \ ref {ChinaOri}, there is a jump in those data, so it was decided to correct that jump by putting the data of the day 02/13/2020 equal to the previous day and that of 02/14/2020 equal to the following day.
With these corrections, the following data and the graph in figure \ref{ChinaCorr} are obtained.

$$
\begin{array}{ccccc}
fecha &  	M &  	a &  	b &  		\alpha \\\hline 
31/03/2020	 &  66871 &  	-0.15032863 &  	4.21103499 &  		4.436336 \\\hline
\end{array}
$$
In this case, a $ R^2 = 0.999885 $ is obtained, which is better than the data without correction.

Note that the latest data shows that there is a linear trend, this can be improved as seen in the case of South Korea, as seen in the following subsection.

\subsection{South Korea: original data}
\label{CoreaO}
\begin{figure}[ht]
\centerline{\includegraphics[width=9cm,height=6cm]{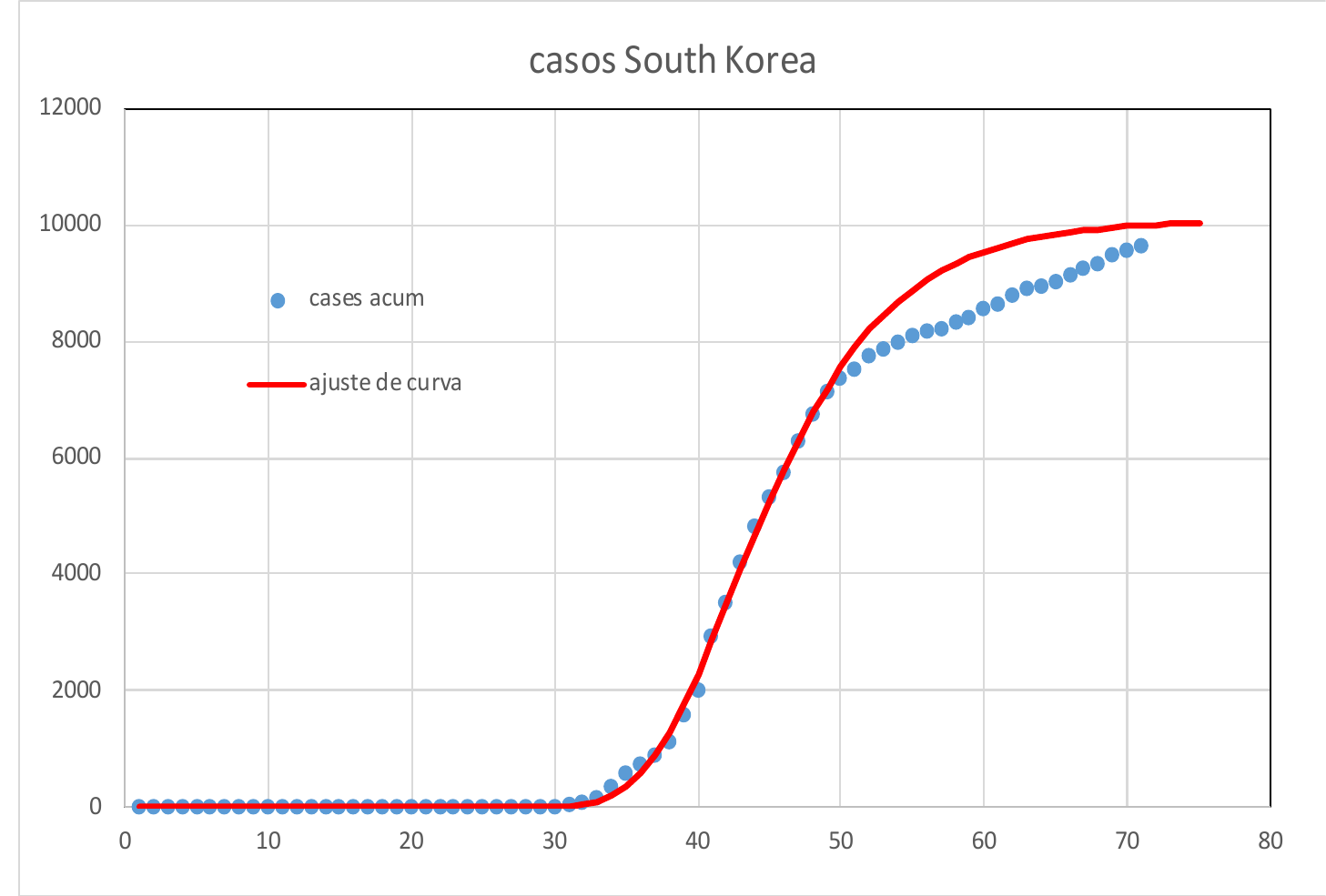}}
\caption{Fit with GLR,  South Korea data}
\label{CoreaSur}
\end{figure}

Applying the method to the data for South Korea gives the following data and the graph in figure \ref{CoreaSur}.
$$
\begin{array}{ccccc}
fecha &  	M &  	a &  	b &  		\alpha \\\hline 
31/03/2020	 &  10079 &  	-0.16416438 &  	0.18353146 &  		874.692503   \\\hline
\end{array}
$$
And we get a $ R ^ 2 = 0.99875 $, which is ``not very good'' as seen in figure \ref{CoreaSur} that the fit is ``not very good''.

To improve this fit we propose to use the fit curve (\ref {logGenrecta}), that is:

$$
P(t) \ = \ \frac{M + ct}{\big(1+e^{-at+b}\big)^{\alpha}}.
$$
This curve improves the fit by obtaining the following data:
$$
\begin{array}{cccccc}
fecha &  	c &  	M &  	a &  	b &  		\alpha \\\hline 
31/03/2020	 &  92.63407116 &  	3046 &  	-0.36922771 &  	15.5617475 &  		0.9691940884 \\\hline
\end{array}
$$

\begin{figure}[ht]
\centerline{\includegraphics[width=9cm,height=6cm]{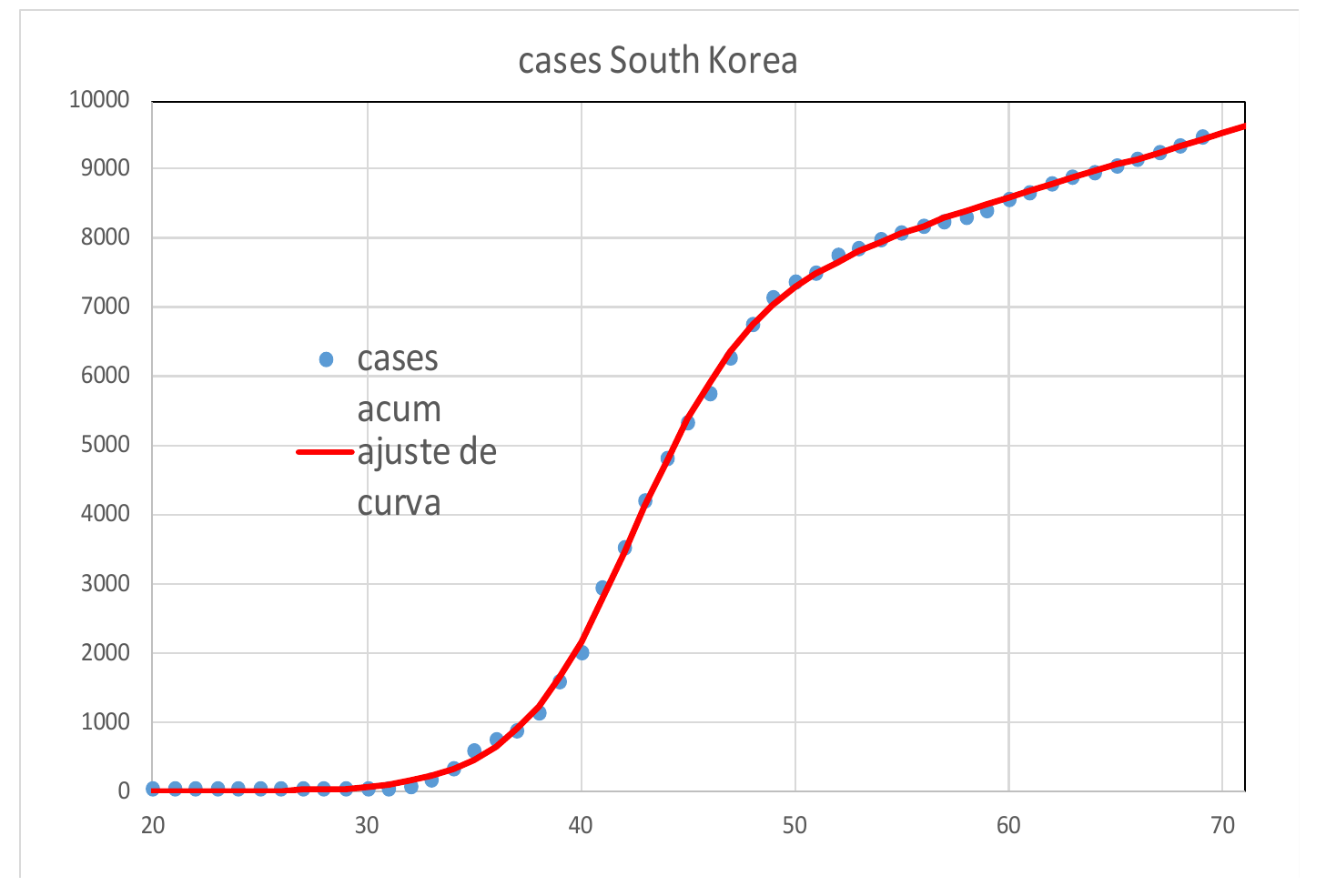}}
\caption{fit with GLR, infected in South Korea {\bf with linear final trend}.}
\label{CoreaSurRecta}
\end{figure}

And now the $ R^2 = 0.99988 $ which is better than the previous result and as seen in the graph in figure \ref{CoreaSurRecta}.
	
As seen, the fit with this type of generalized logistic curves is very good, obtaining $ R ^ 2 $ greater than 0.9999.

\newpage
\section{Prediction tests}
\label{PbaPred}

As we saw in the previous section, the logistic curves approximate very well, as has been shown in previous works (for example \cite{Fekedulegn, Pella}).

In this section we will try to measure the relative error percentage that is made when using the proposed curves, a test curve will be used and the data from China and South Korea

\subsection*{Test curve}

To try to validate the hypothesis we are going to generate a curve with values
$$
\begin{array}{cccc}
 	M &  	a &  	b &  		\alpha \\  \hline
  2000 &  	-0.05 &  	-2&  		10 \\\hline  
\end{array}
$$

and when using the optimization algorithm, the following upper bounds are obtained for the relative error
When using the LG function, with the indicated number of days:

\begin{center}
{\sc Relative error percentage in test curve using GLR}

\begin{tabular}{llr}
number of days & max rel. error\\\hline
25 	&	10.451\% \\	
30 	&	4.801\%	\\
35 	&	1.825\%\\
40 	&	1.188\%	\\\hline
\end{tabular}
\end{center}

Using the Gompertz function improves the prediction, as mentioned by having fewer parameters, when using the example with
$$
\begin{array}{cccc}
 	a &  	b &  		c \\ \hline 
  2000 &   	-2&  		10   
\end{array}
$$
When executing the method the following relative error percentages are obtained

\begin{center}
{\sc Relative error percentage on test curve \\ using Gompertz}

\begin{tabular}{cc}
number of days & max rel. error\\\hline
25 	&		1.6697\% \\	
30 	&		0.7492\%	\\
35 	&	0.8928\%\\
40 		&	0.3234\%	\\\hline
\end{tabular}
\end{center}

As seen with the Gompertz function, it seems that better predictions can be guaranteed.

\newpage
\subsection*{China Data}

Let us now look at the case of China, where the curve is almost complete.

As we see when using the days from 16 to 45 (20 days), note that the first 16 days there were almost no changes in the data, so that data was not used, with these data with GLR we obtain:

\subsection*{Gompertz}

When using the days from 16 to 45 (20 days) a maximum error of 9.91 \% is obtained (average of 3 runs, basically gives the same value)

\begin{figure}[ht]
\centerline{\includegraphics[width=15cm,height=9cm]{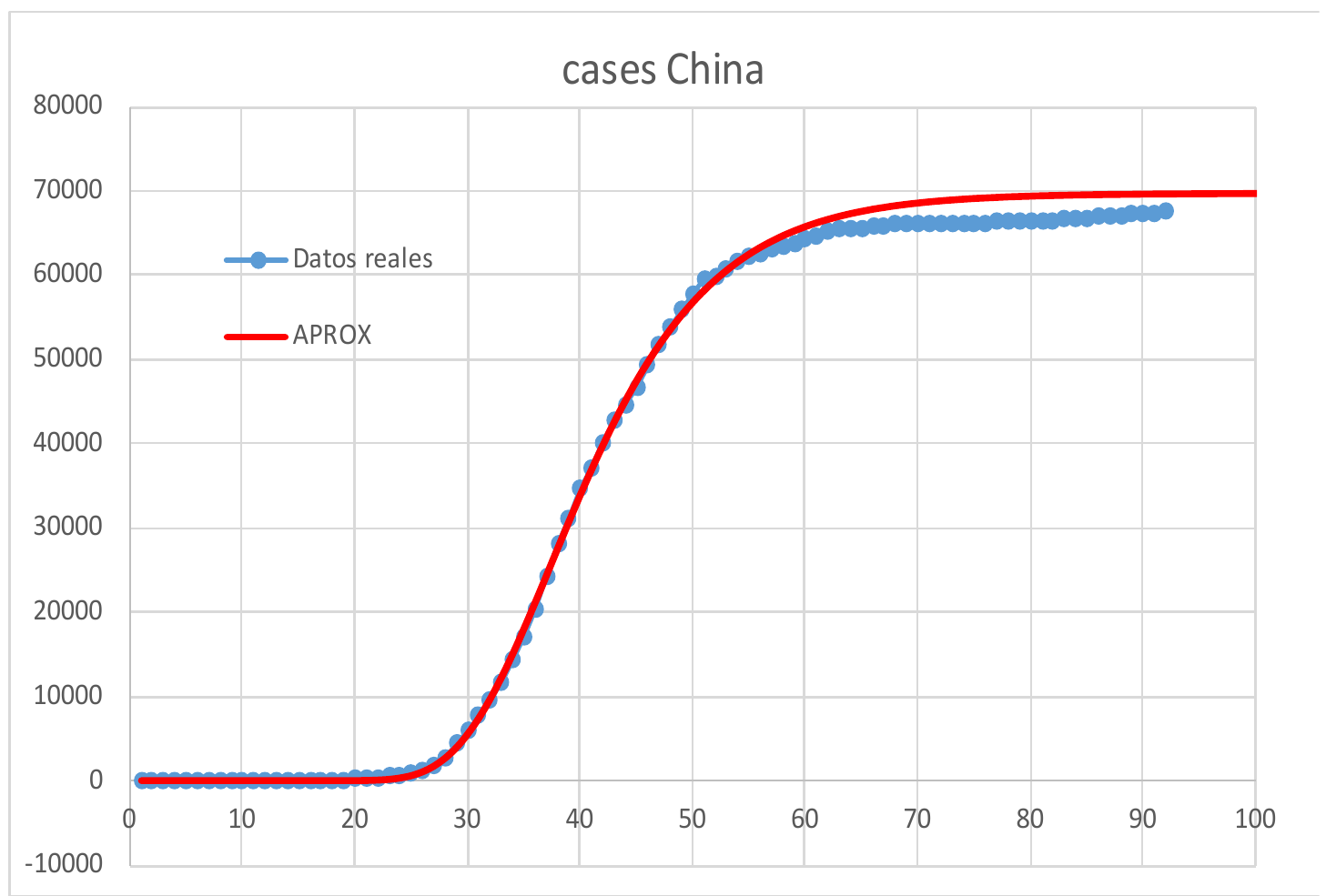}}
\caption{30 day forecast with Gompertz, China }
\label{predChinaG}
\end{figure}

When using the days from 15 to 55 (30 days), the graph of figure \ref {predChinaG} is obtained. When measuring the relative errors, a maximum value of 4.49 \% (average of 3 runs) is obtained, in the prediction of the data until 03/31/2020.

\begin{center}
{\sc Relative error percentage prediction China \\ using Gompertz}

\begin{tabular}{cc}
number of days & max rel. error\\\hline
20 	&		9.91\% \\	
25 	&		4.49\%	\\\hline
\end{tabular}
\end{center}

\subsection*{South Korea Data}

\begin{figure}[ht]
\centerline{\includegraphics[width=15cm,height=9cm]{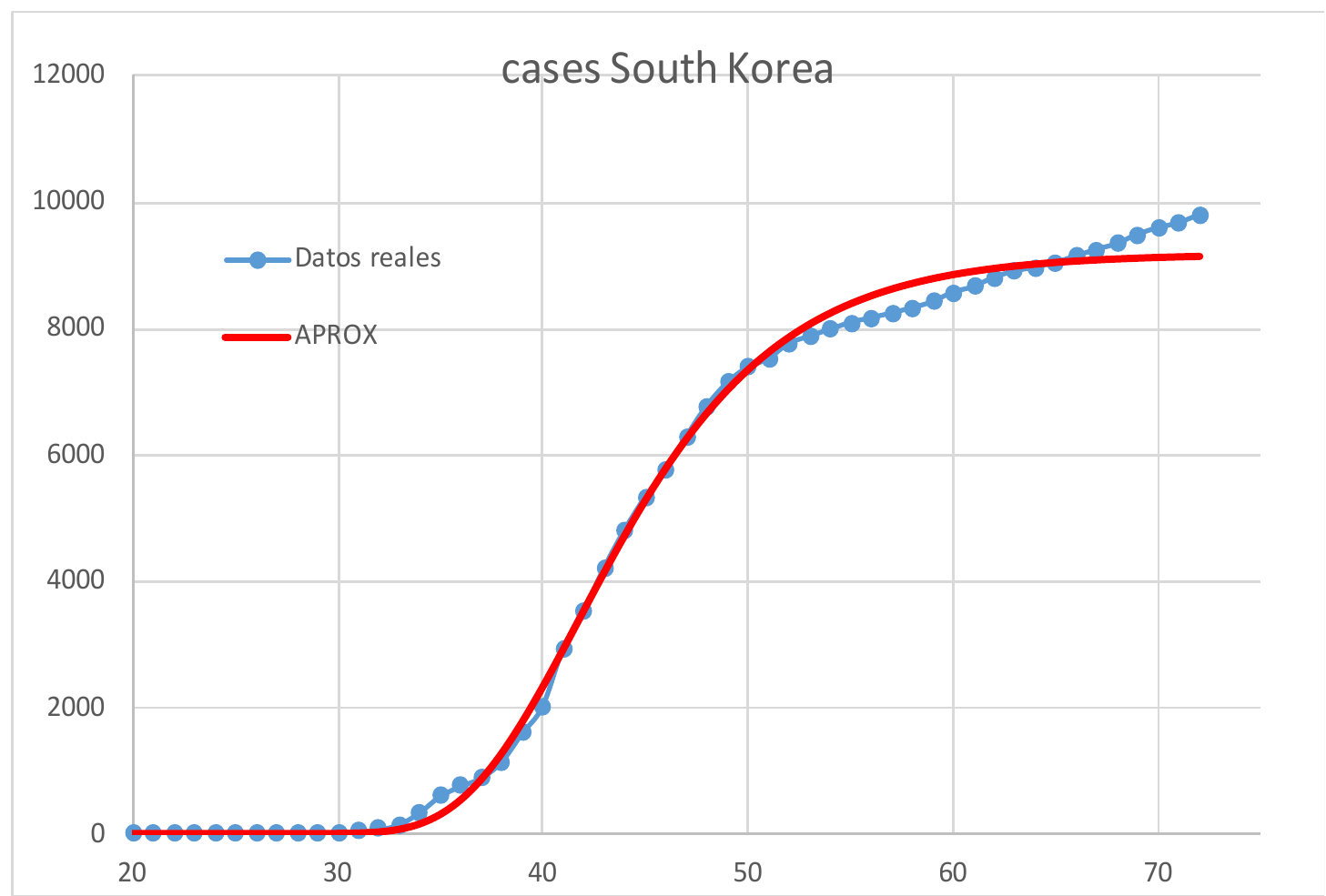}}
\caption{25 days forecast with Gompertz, South Korea, }
\label{predCoreaG}
\end{figure}

For South Korea we will use the data from day 28, since the first case (16/02/202), in this case remember that this data has a linear trend in last days, see figure \ref{predCoreaG} .

\begin{center}
{\sc Relative error percentage in South Korea \\ prediction using Gompertz}
\medskip

\begin{tabular}{cc}
 number of days  &  máx error rel.\\\hline
 20 	&		17.83\% \\	
 25 	&		6.49\%	\\\hline
\end{tabular}
\end{center}

% \end{document}

\section{Results}

As results of this work and as we have seen, this method will be used to give a prediction for the data of some countries starting with Costa Rica and then the data for Italy, Spain will be presented, and with the adjustments that we have, we already have that of China. and South Korea.

\subsection{China and South Korea}

In the case of China and South Korea, the results obtained in the subsections \ref{ChinaO} and \ref{CoreaO} can be used.

%\newpage
\subsection{Costa Rica}

For Costa Rica, the following data is available from  March 6, when the first case was detected.
\begin{center}
{\sc Covid-19 data in Costa Rica}\bigskip

\begin{tabular}{|lcc||lcc||lcc|}\hline
&	daily &	Total & 	&	daily &	Total & 	&	daily &	Total \\
Date &  cases & 	  cases &	Date & cases & 	  cases &	Date & cases & 	  cases \\\hline
06/03/20 & 	1 & 	1 & 	15/03/20 & 	8 & 	35 & 	24/03/20 & 	19 & 	177 \\ 
07/03/20 & 	4 & 	5 & 	16/03/20 & 	6 & 	41 & 	25/03/20 & 	24 & 	201 \\ 
08/03/20 & 	4 & 	9 & 	17/03/20 & 	9 & 	50 & 	26/03/20 & 	30 & 	231 \\ 
09/03/20 & 	4 & 	13 & 	18/03/20 & 	19 & 	69 & 	27/03/20 & 	32 & 	263 \\ 
10/03/20 & 	4 & 	17 & 	19/03/20 & 	18 & 	87 & 	28/03/20 & 	32 & 	295 \\ 
11/03/20 & 	5 & 	22 & 	20/03/20 & 	26 & 	113 & 	29/03/20 & 	19 & 	314 \\ 
12/03/20 & 	1 & 	23 & 	21/03/20 & 	4 & 	117 & 	30/03/20 & 	16 & 	330 \\ 
13/03/20 & 	3 & 	26 & 	22/03/20 & 	17 & 	134 & 	31/03/20 & 	17 & 	347 \\ 
14/03/20 & 	1 & 	27 & 	23/03/20 & 	24 & 	158 & 	01/04/20 & 	28 & 	375 \\ 
06/03/20 & 	1 & 	1 & 	15/03/20 & 	8 & 	35 & 	24/03/20 & 	19 & 	177 \\ 
07/03/20 & 	4 & 	5 & 	16/03/20 & 	6 & 	41 & 	25/03/20 & 	24 & 	201 \\ 
08/03/20 & 	4 & 	9 & 	17/03/20 & 	9 & 	50 & 	26/03/20 & 	30 & 	231 \\ 
09/03/20 & 	4 & 	13 & 	18/03/20 & 	19 & 	69 & 	27/03/20 & 	32 & 	263 \\ 
10/03/20 & 	4 & 	17 & 	19/03/20 & 	18 & 	87 & 	28/03/20 & 	32 & 	295 \\ 
11/03/20 & 	5 & 	22 & 	20/03/20 & 	26 & 	113 & 	29/03/20 & 	19 & 	314 \\ 
12/03/20 & 	1 & 	23 & 	21/03/20 & 	4 & 	117 & 	30/03/20 & 	16 & 	330 \\ 
13/03/20 & 	3 & 	26 & 	22/03/20 & 	17 & 	134 & 	31/03/20 & 	17 & 	347 \\ 
14/03/20 & 	1 & 	27 & 	23/03/20 & 	24 & 	158 & 	01/04/20 & 	28 & 	375 \\ \hline

\end{tabular}

\end{center}

With these data when making the adjustment with the GL function, the following parameters are obtained, and with this amount of data, basically there is only one optimum:

\begin{center}
{\sc Parameters obtained for Costa Rica \\ using Generalized Logistics}

\begin{tabular}{cccc}
M & 	a & 	b & 		c \\\hline 
886 & 	-0.0780135 & 	-4.91608521 & 		1023.90189 \\ \hline
\end{tabular}
\end{center}
This curve fit has a $R^2 =	0.99858343$.

\begin{figure}[ht]
\centerline{\includegraphics[width=15cm,height=9cm]{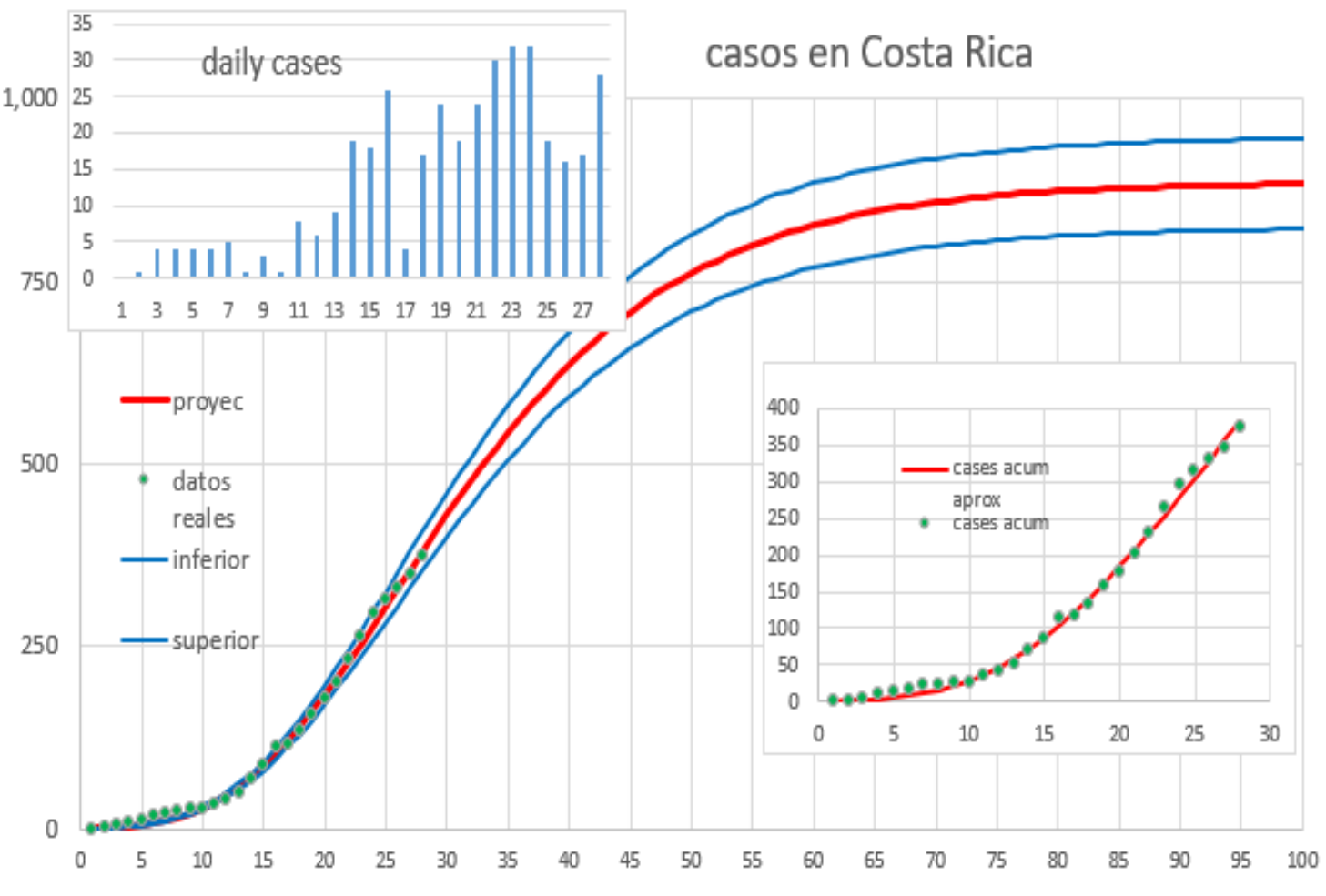}}
\caption{Forecast with LG, Costa Rica}
\label{CostaRica}
\end{figure}

In the graph in figure \ref{CostaRica} you can see the prediction of cases, you can see the number of daily cases in the upper box and the adjustment of the real data in the lower box.  

Upper and lower bounds are included for the forecast, with an error of 7 \% in the data, base on the results in the section \ref{PbaPred}.

It is also seen that if the situations are as they are today there will be a limit of approximately 1000 cases, that is, from the mathematical point of view or curve fitting.\bigskip
 
{\bf Gompertz}\bigskip

With the Gompertz curve, basically the same result is obtained as seen in the following:
\bigskip

\begin{center}
{\sc Parameters obtained for Costa Rica \\ using Gompertz}

\begin{tabular}{cccc}
 	a & 	b & 		c \\\hline 
887 & 	-6.923348908 & 		-0.077891632 \\ 
\end{tabular}
\end{center}
\bigskip

\begin{figure}[ht]
\centerline{\includegraphics[width=15cm,height=9cm]{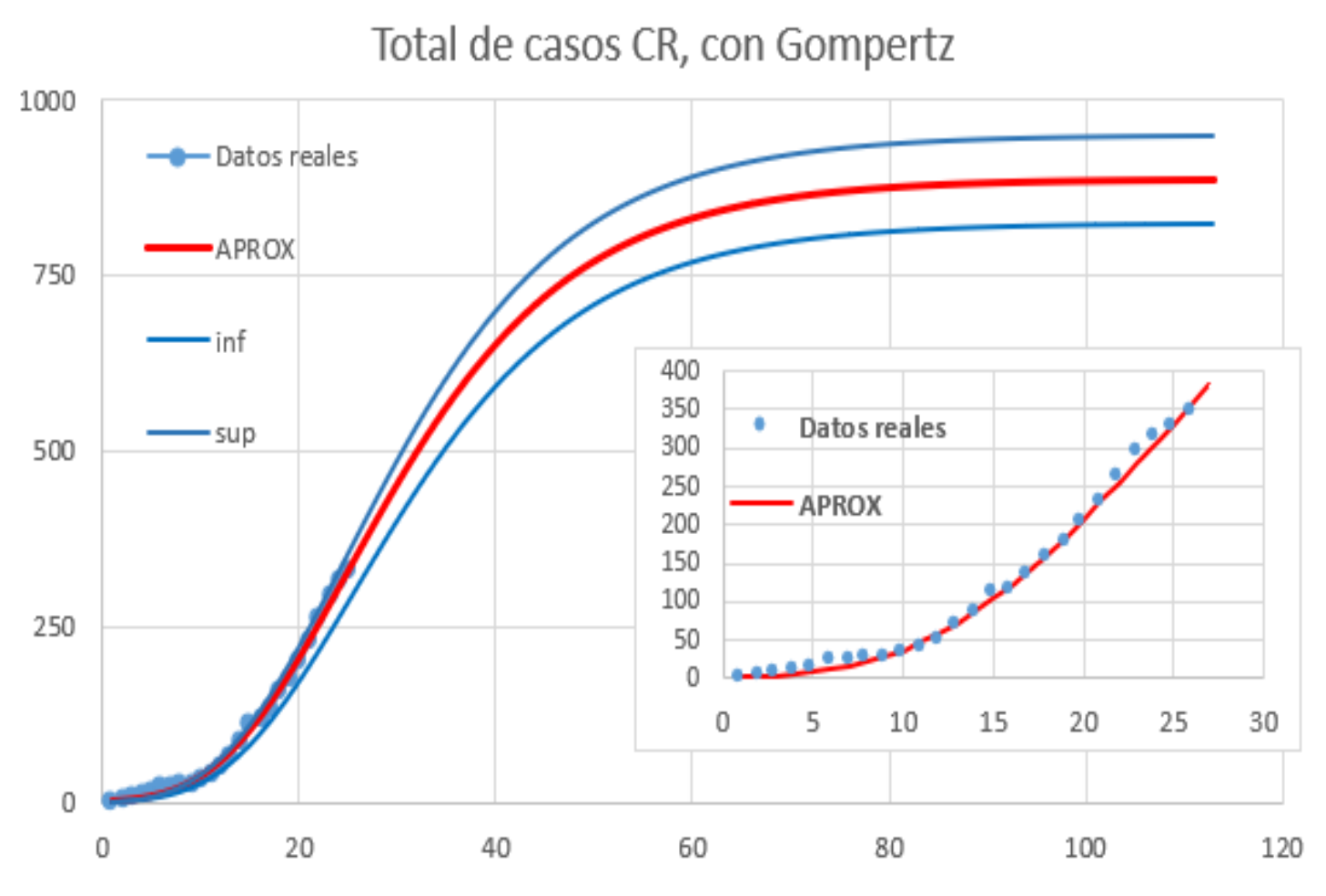}}
\caption{Forecast with Gompertz, Costa Rica}
\label{CostaRicaG}
\end{figure}

This fit has a $ R ^ 2 = 0.99853504 $, in the figure \ref{CostaRicaG} you can see the resulting graph.

It should be noted that the results of using these 2 functions have been approximating each other as the days go by and the maximum value between them is presented in the following table for the sample.

Furthermore, it can be seen that as the days go by, the limit value of $ M $ decreases, possibly due to the measures taken by the country's Government.

\newpage
\begin{center}
{\sc Limit value of $ M $ of the forecast according\\ to day,  Costa Rica}
\\
\begin{tabular}{cccc}
	Date &	Gompertz &	LG \\\hline
23/03/20 & 	2162 & 	1983 \\ 
24/03/20 & 	1445 & 	1359 \\ 
25/03/20 & 	1310 & 	1274 \\ 
26/03/20 & 	1583 & 	1505 \\ 
27/03/20 & 	2048 & 	1974 \\ 
28/03/20 & 	2426 & 	2343 \\ 
29/03/20 & 	1743 & 	1730 \\ 
30/03/20 & 	1193 & 	1185 \\ 
31/03/20 & 	924 & 	922 \\ 
01/04/20 & 	887 & 	886 \\ 
02/04/20 & 	829 & 	828 \\ 
03/04/20 & 	788 & 	789 \\ 
04/04/20 & 	761 & 	761 \\ 
05/04/20 & 	744 & 	743 \\ 
\hline
\end{tabular}
\end{center}

\begin{figure}[ht]
\centerline{\includegraphics[width=11cm,height=6cm]{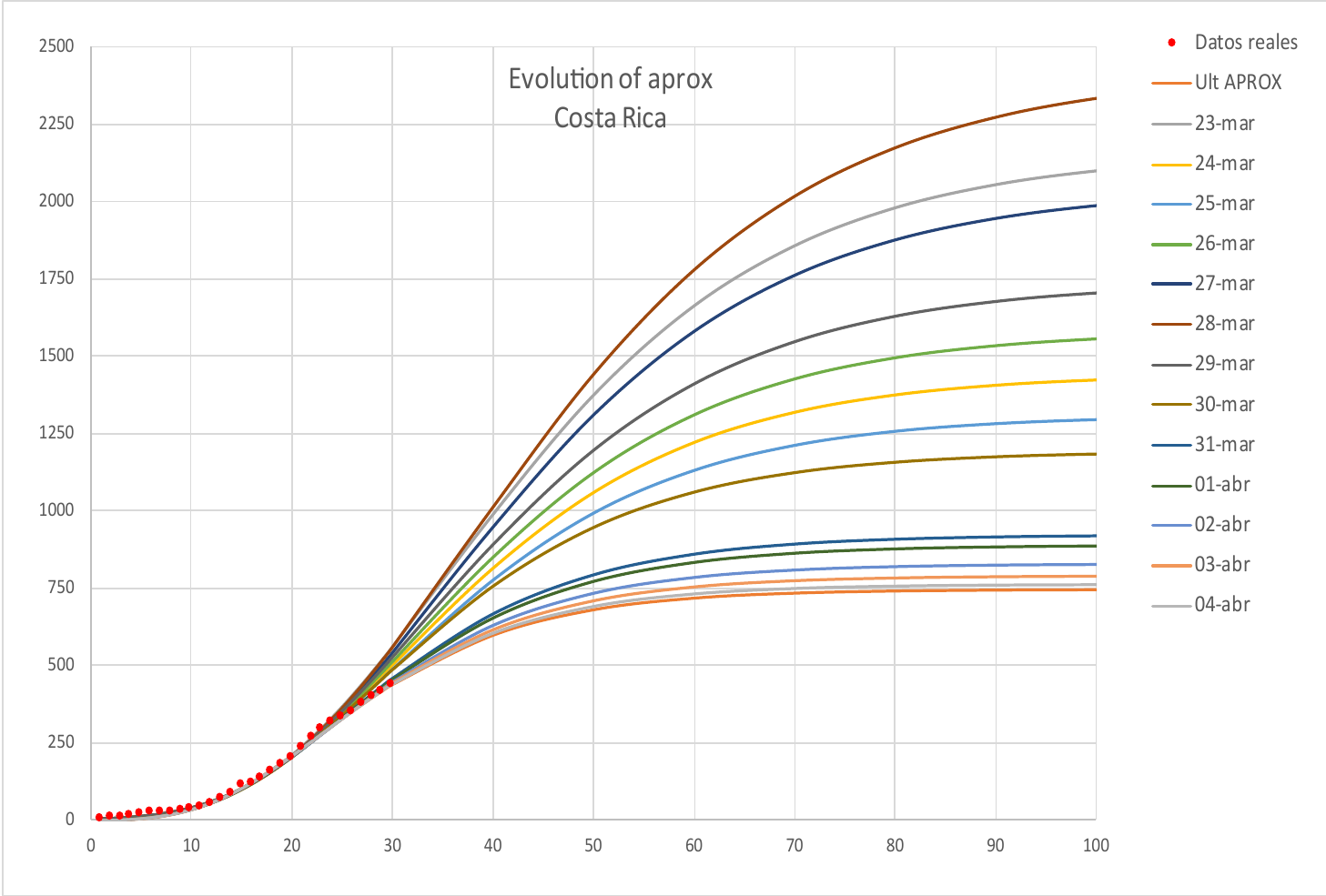}}
\caption{Costa Rica, forecast evolution}
\label{CostaRicaevol}
\end{figure}

Furthermore, it can be seen that as the days go by, the limit value of $ M $ decreases, possibly due to the measures taken by the country's Government, in the figure \ref{CostaRicaevol} this is show. 
\bigskip
%\vspace{-13em}

\newpage
\

\newpage
\subsection{Italy}

For Italy, the results obtained using Gompertz and with the data as of 3/30/2020 are presented.

\begin{figure}[ht]
\centerline{\includegraphics[width=15cm,height=9cm]{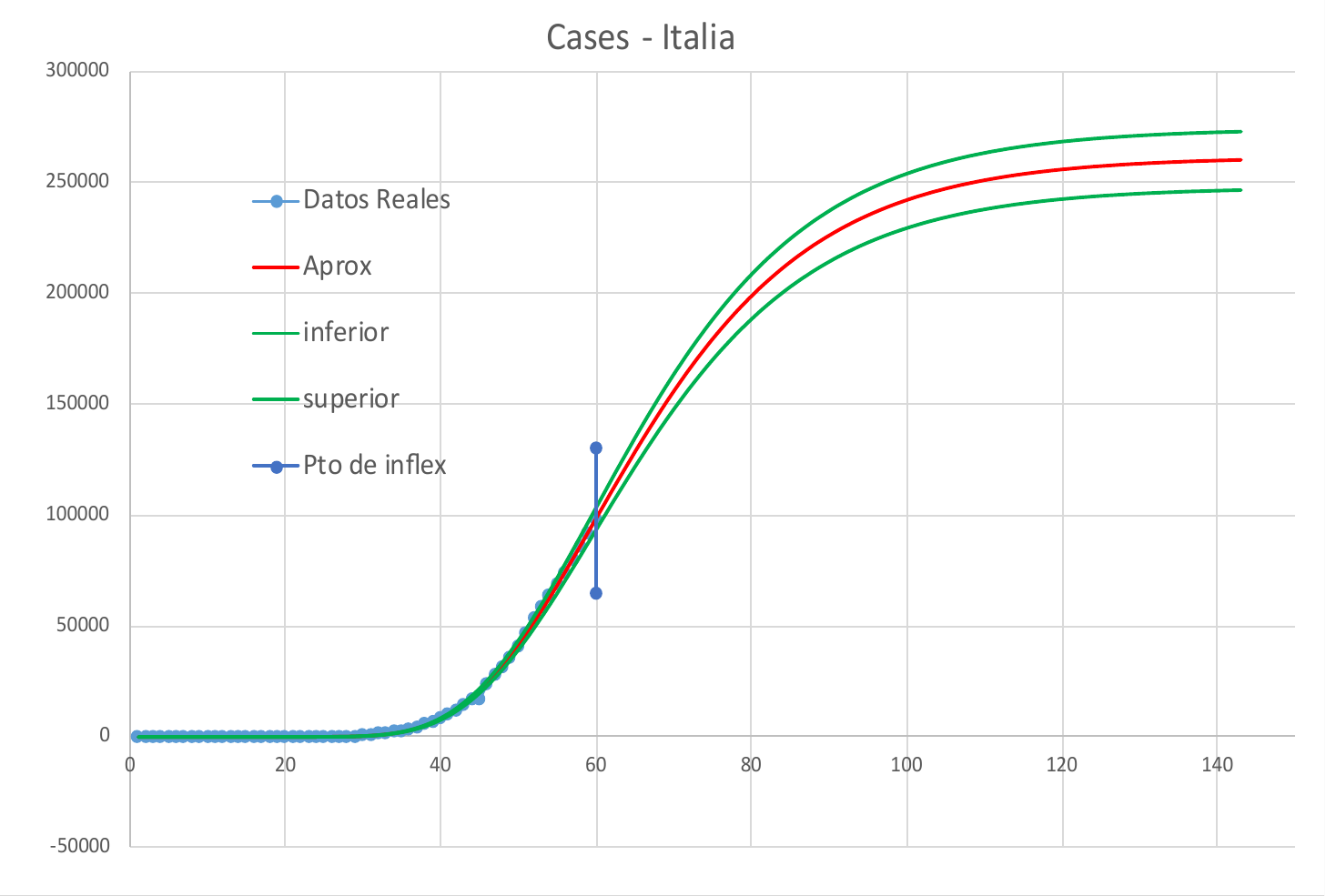}}
\caption{Italy, forecast with Gompertz,}
\label{Italia}
\end{figure}

\begin{center}
{\sc Parameters obtained for Italy \\ using Gompertz}
 
\begin{tabular}{cccc}
 	a & 	b & 		c \\\hline 
261 052 & 	-43.77482253 & 		-0.063450536 \\ 
\end{tabular}
\end{center}
This fit has a $ R^2 = 0.99968 $, in the figure \ref{Italia} you can see the resulting graph.

\newpage

\subsection{Spain}

For Spain, the results are presented when using the Gompertz function and with the data as of 3/31/2020.

\begin{figure}[ht]
\centerline{\includegraphics[width=15cm,height=9cm]{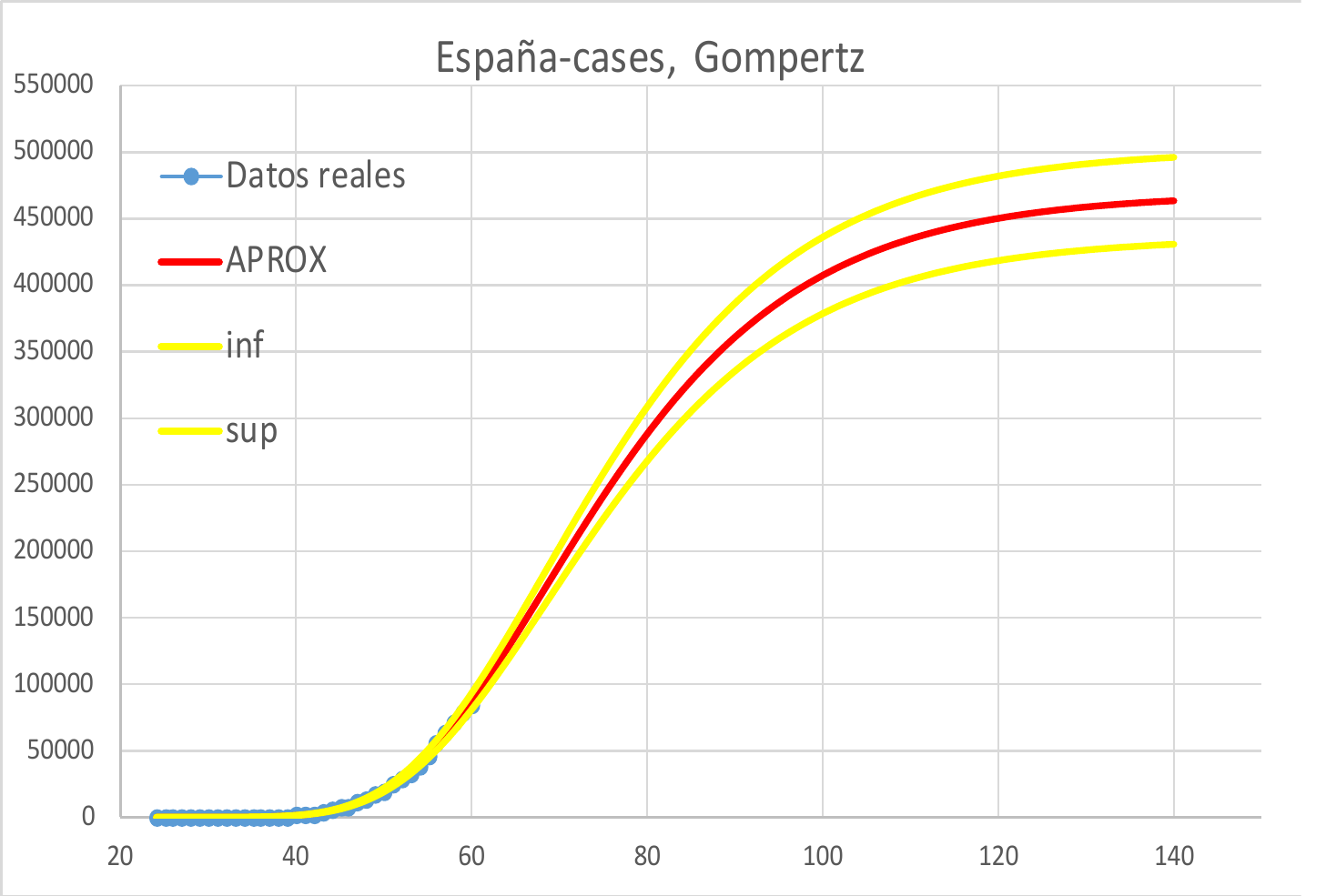}}
\caption{Spain, Forecast with Gompertz.}
\label{Espana}
\end{figure}

\begin{center}
{\sc Parameters obtained for Spain \\ using Gompertz}
 
\begin{tabular}{cccc}
 	a & 	b & 		c \\\hline 
468 495 & 	-69.20043418 & 		-0.061995389 \\ 
\end{tabular}
\end{center}
This fit has a $ R^2 = 0.999410067 $, in the figure \ref{Espana} you can see the resulting graph. 

\newpage

\subsection{USA}

For USA, the results are presented when using the Gompertz function and with the data as of april 5,2020.

\begin{figure}[ht]
\centerline{\includegraphics[width=15cm,height=9cm]{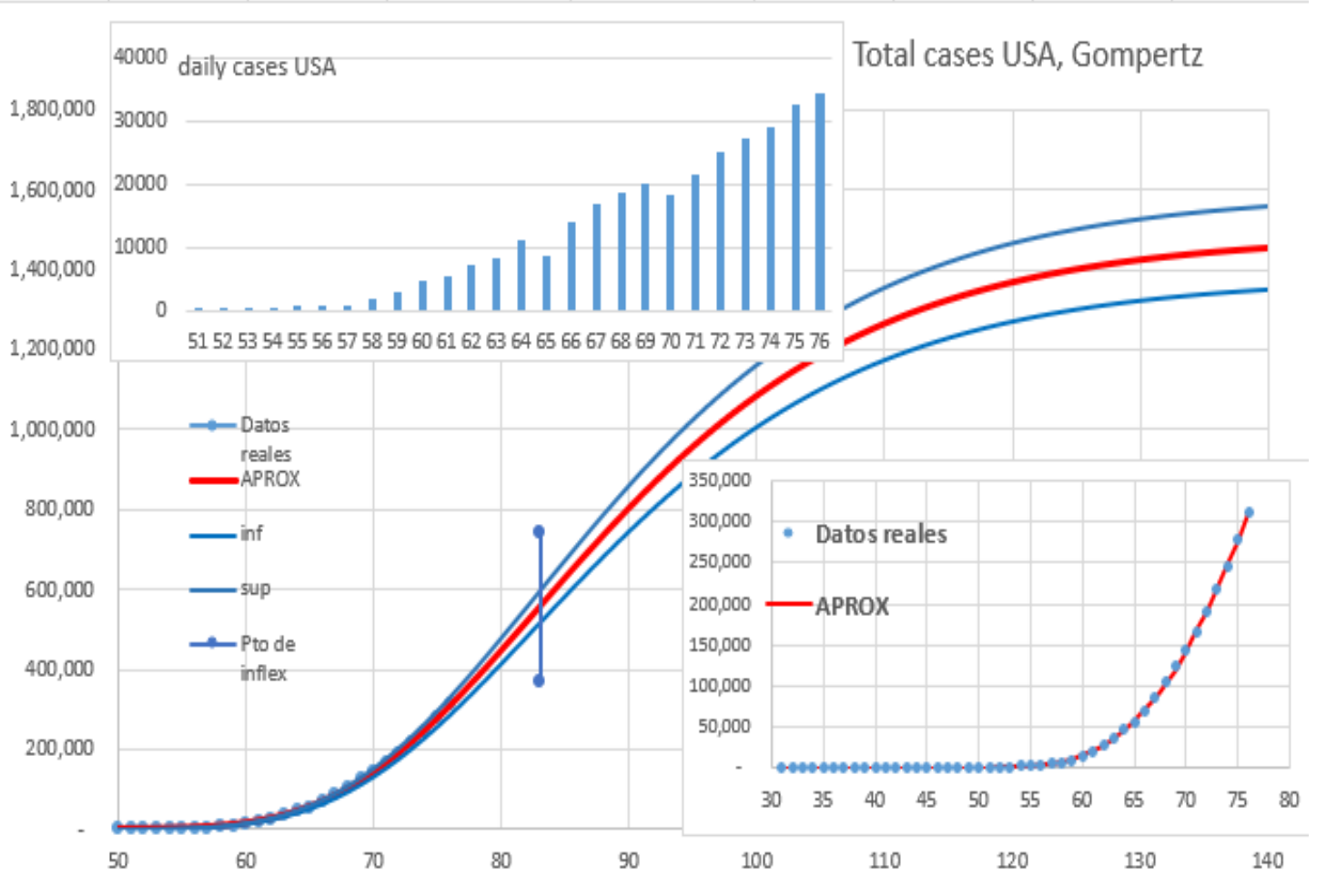}}
\caption{USA, Forecast with Gompertz.}
\label{USA}
\end{figure}

\begin{center}
{\sc Parameters obtained for USA \\ using Gompertz}

\begin{tabular}{cccc}
 	a & 	b & 		c \\\hline 
 1 486 347  & 	-253.23772 & 		-0.066899181 \\ 
\end{tabular}
\end{center}
This fit has a $ R^2 = 0.999915 $, in the figure \ref{USA} you can see the resulting graph. 

\subsection{Other countries}

There are the results of other countries that will be incorporated later

\section{Conclusions and future work}

This work shows how population growth curves can be adjusted, using the LG and Gompertz functions, even in the most general case, such as the data for South Korea.
section \ref{CoreaO}, where logistics is being carried out with a straight line.

It is seen that these methods could be used to predict the growth of populations, in this case of people infected with Covid-19, and could help pandemic experts to take the necessary measures.

\subsection{Future work}

More studies are needed to refine the results of this work.

See the possibility of using similar curves to approximate other types of data.

\end{document}